\documentclass[preprint]{elsarticle}

\usepackage{booktabs, ctable, dcolumn}
\usepackage{epsfig, amsmath}
\usepackage{datetime}
\usepackage{graphicx}    
\usepackage{graphicx}
\usepackage{ulem}
 \usepackage{multirow}

\newcommand{\be}{\begin{equation}}
\newcommand{\ee}{\end{equation}}
\newcommand{\bea}{\begin{eqnarray}}
\newcommand{\eea}{\end{eqnarray}}
\newcommand{\uv}[1]{\ensuremath{\mathbf{\hat{#1}}}} 
\newcommand{\tnsr}[1]{\overset\leftrightarrow{#1}} 

\newcommand{\Rmnum}[1]{\expandafter\@slowromancap\romannumeral #1@}
\newcommand{\tab}{\hspace{6mm}}
\begin{document}

\begin{frontmatter}

\title{A surface-scattering model satisfying energy conservation and reciprocity}
\author[label1]{Karthik Sasihithlu \corref{cor1}\fnref{fn1} }
\author[label2]{Nir Dahan}
\author[label1]{Jean-Paul Hugonin}
\author[label1]{Jean-Jacques Greffet}

\address[label1]{Laboratoire Charles Fabry, Institut d'Optique, CNRS - Universit\'e Paris-Sud, Campus Polytechnique, RD128, 91127, Palaiseau Cedex, France}

\address[label2]{SCD SemiConductor Devices P.O.Box 2250, Haifa 31021, Israel}

\cortext[cor1]{k.sasihithlu@imperial.ac.uk}

\fntext[fn1]{Present address: The Blackett laboratory, Imperial College London, London SW7 2AZ, UK}

\date{\ddmmyyyydate \today}

\begin{abstract}
In order for surface scattering models to be accurate they must necessarily satisfy energy conservation and reciprocity principles. Roughness scattering models based on Kirchoff's approximation or perturbation theory do not satisfy these criteria in all frequency ranges. Here we present a surface scattering model based on analysis of scattering from a layer of particles on top of a substrate in the dipole approximation which satisfies both energy conservation and reciprocity and is thus accurate in all frequency ranges. The model takes into account the absorption in the substrate induced by the particles but does not take into account the near-field interactions between the particles. 
\end{abstract}


\end{frontmatter}

\section{ \label{intro} INTRODUCTION}
%
%
%

%


In this paper, we report a model for scattering at an interface separating two homogeneous media. The aim is to derive a BSDF (Bidirectional Scattering Distribution Function) that accounts for both specular and diffuse components of light scattered both in transmission and in reflection, and also absorption. The initial motivation of this work is to develop a model that can be used to analyze multiple scattering in a photovoltaic cell. Such a cell is a multi-layer system comprising an absorbing active layer (e.g. silicon or CiGS) deposited on different materials that can be used as electrical contacts, back-reflectors, antireflection coatings, etc. Most active materials used for photovoltaic cells are semiconductors whose refractive index takes large values so that accounting for total internal reflection is very important. A basic mechanism that is often used to increase the absorption  is to scatter light in order to trap the light in the absorbing medium. Most designs are performed either using a periodic model for the roughness or by trial and error  using randomly rough surfaces.

\begin{figure}[Hhbt]
\begin{center}
\includegraphics[trim = 0cm 0cm 0cm 0cm, clip=true, scale=0.4]{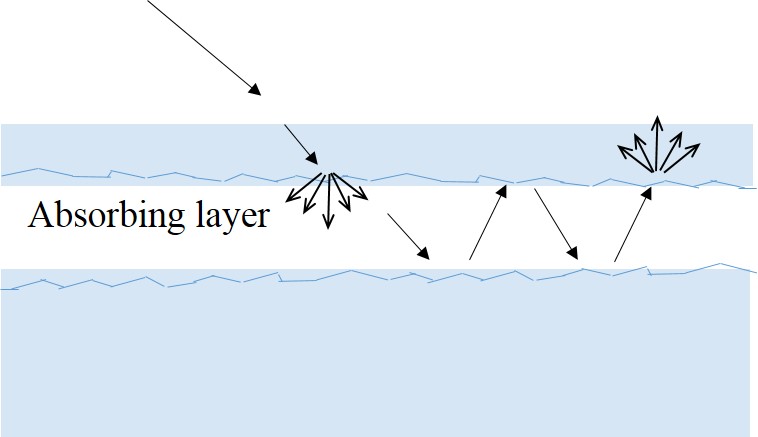}
\caption{Light trapped in a layer can undergo both coherent reflection and transmission as well as multiple scattering}
\label{Intro}
\end{center}
\end{figure}

In this paper, we explore the issues raised by the modeling of scattering and absorption in such a structure using a radiative transfer equation approach. For a multi-layer geometry, an adding doubling approach to the solution of the radiative transfer equation is appropriate. Hence, the basic tool needed, is a matrix accounting for the BSDF of the interface. The key issue as far as photovoltaic applications are concerned is that the BSDF needs to fulfil energy conservation with an accuracy better than 1\%. Indeed, the whole point of the modeling of a photovoltaic cell is to gain a few percent in the absorption. This sets the standards required for the accuracy of the model. Not less important is the requirement of reciprocity. Indeed, when a cell is designed in order to scatter light to couple incident propagating light to trapped light in the cell, the same roughness can couple back trapped light to propagating modes with the same efficiency owing to reciprocity. Hence, if light can be coupled to guided modes, then guided modes can be coupled back to propagating modes and escape as shown in Fig. \ref{Intro}. Thus, optimizing the right coupling between guided modes and propagating modes is far from trivial and requires an accurate model that accounts properly for reciprocity.

Currently available models \cite{schroder2011modeling, hermansson2003review, elfouhaily2004critical} for scattering by rough surfaces are focussed on modeling properly the angular scattering pattern and describing properly the diffuse and collimated reflection factors. Satisfying reciprocity and energy conservation is not the major issue for most of them. As a matter of fact, models based on Kirchhoff approximation, phase perturbation, etc do not satisfy energy conservation and reciprocity with good accuracy over all frequency ranges. For energy applications, what is needed is not just an exact angular description of the scattering but also a correct modeling of the balance between collimated and diffuse scattering as well as a correct balance between scattering and absorption. When dealing with light propagating in scattering media such as gases, particles or paper for example, the isotropic approximation for the phase function is very often used \cite{chandrasekhar2013radiative}.  Here, we  establish a similar approximation while accounting for the presence of interfaces. Inasmuch as the system is in a multiple scattering regime, the final result does not depend much on the details of the phase function. Instead, accounting properly for absorption and scattering and satisfying energy conservation is critical. Hence, approximations such as Milne-Eddington, diffusion approximation, \cite{thomas2002radiative, mishchenko2002scattering} etc are extremely useful. This discussion sets the landscape: we seek a model that allows introducing scattering and absorption with enough parameters to control the balance between collimated and diffuse, and between scattering and absorption. As we are looking for a multiple scattering regime, the accuracy in modeling the exact angular behavior is not very important.  By contrast, it is of critical importance that the model satisfies energy conservation and reciprocity.

In order to establish such a model, we start from the scattering of a single dipolar scatterer located at a certain distance from an interface. For such a scatterer characterized by its polarizability $\alpha$ such that its dipole moment is proportional to  
$\alpha$ and the incident field on the scatterer,  we can derive explicitly the scattered field in an electrodynamics framework. This solution satisfies energy conservation and reciprocity. It does also account for the coupling of light into guided modes in the medium with high refractive index. Finally, it also accounts for absorption induced in lossy substrates by the near-field produced by the scatterer. This elementary result, which was extensively discussed in Ref.\cite{dahan2012enhanced} will serve as a building block for our model. In this paper, we take a further step in order to develop a model for the scattering. We need to establish a model for the coherent modification of the reflection and transmission factor for the collimated beams as well as account for diffuse scattering and absorption. The available model parameters are as follows: the real and imaginary part of the polarizability, the number of scatterers per unit surface, and the distance between the scatterer and the interface. In the low density regime, our model will be expected to fit with an exact numerical solution of the problem of scattering by N scatterers. For  larger density, near-field and correlation effects are expected to become important. Our goal in this paper is not to capture these effects accurately. Instead, we only look for a model that allows introducing a degree of scattering while preserving energy conservation and reciprocity.

The paper is arranged as follows:  In Sec. \ref{sec:single} a brief description of scattering from a particle in a homogeneous medium is given which introduces the terminology used in this paper. This is then extended to the case of scattering from a layer of particles in a homogeneous medium in Sec. \ref{sec:eff_index} where we use the effective index method based on Maxwell Garnet theory to analyze the coherent scattering from the particles.  We then use this theory in Sec. \ref{sec:layer_substrate} to obtain the BSDF of a surface with a layer of particles on top of it serving as source for the scattering function and compare it in Sec. \ref{exact} with exact numerical simulation of scattering from a layer of particles on top of a finite surface. Finally, in  \ref{sec:microscopic} we show the microscopic  derivation of coherent scattering from a layer of particles in a homogeneous medium where we take into account the multiple scattering between the particles using the mean-field theory but ignore near-field interactions and correlations (and hence recurrent scattering) between them, thereby establishing the equivalence between the effective-index model and coherent scattering in the mean-field approximation.

\section{\label{sec:single}Scattering from a single particle in a homogeneous medium}

\begin{figure}[Hhbt]
\begin{center}
\includegraphics[trim = 0cm 0cm 0cm 0cm, clip=true, scale=0.5]{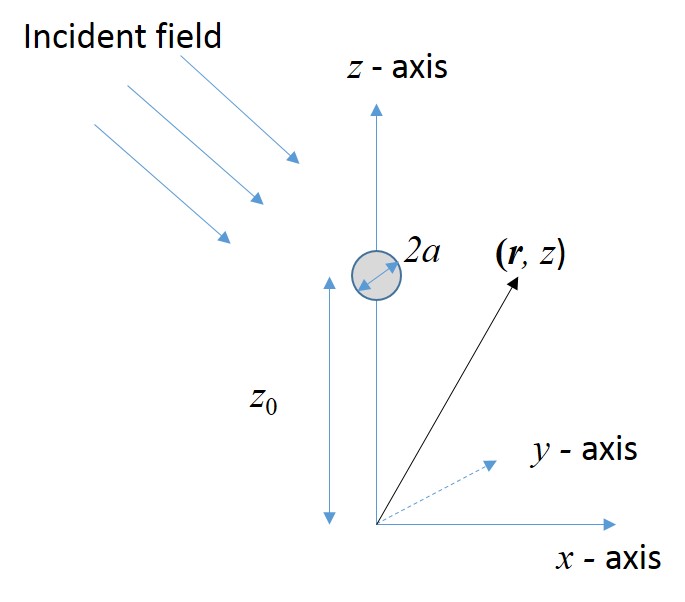}
\caption{Configuration for finding the scattered field at any point $(\boldsymbol{r}, z)$ from a particle with center located at $(0, z_0)$ in a homogeneous medium of dielectric function $\epsilon_1$.    }
\label{particle}
\end{center}
\end{figure}

In the configuration shown in Fig. \ref{particle} consider a scatterer located at $(0,z_0)$ in a homogeneous medium of dielectric permittivity $\varepsilon_1$. For simplicity of analytical description we assume that the surrounding medium is non-absorptive so that $\varepsilon_1$ is real.  The incident field at position $(\boldsymbol{r}, z)$ is chosen to be a planar wave with amplitude unity at the position of the particle, and with a wavelength much greater than the radius of the particle $a$. The incident field can be written as:
\be
\label{Inc}
\mathbf{E}_i (\boldsymbol{ r}, z) = \uv{e}_i \; e^{i k_{z1} (z_0 - z)} e^{i \mathbf{k}_{\text{inc}}^{||}. \boldsymbol{r}}\,.
\ee
where $\mathbf{k}_{\text{inc}}^{||}$ is a real vector denoting the component of the incident wave vector parallel to the $xy$ plane;  $k_{z1}$ is related to $\mathbf{k}_{\text{inc}}^{||}$ as:
 \be
 k_1^2 =\varepsilon_1 (\omega/c)^2 = k_{z1}^2 + \mathbf{k}_{\text{inc}}^{||2} ;
 \label{kz1}
 \ee
where $\omega$ is the frequency of incident radiation and $c$ is the velocity of light.
  $\uv{e}_i$ is the polarization of the incident wave:
\be
\uv{e}_i =
\begin{cases}
\uv{s} = (0 \uv{x} +  (-1) \uv{y} +  0 \uv{z}) & \tab s /\text{TE} - \;\;  \text{polarization}  \\
\uv{p}_{1-} = (k_{z1} \uv{x} + 0 \uv{y} + \mathbf{k}_{\text{inc}}^{||} \, \uv{z})/k_1 & \tab p /\text{TM} - \;\;  \text{polarization}
\end{cases}
\ee
For a TM polarized wave travelling in the positive $z$ direction we would have $\uv{e}_i= \uv{p}_{1+}$ where $\uv{p}_{1+} = (-k_{z1} \uv{x} + 0 \uv{y} + k_{\text{inc}}^{||} \, \uv{z})/k_1$.
%

The incident field polarizes the particle such that its dipole moment is given by \cite{draine1988discrete}:
\be
\label{p0}
\mathbf{p}_0 = \varepsilon_0 \alpha_0 \mathbf{E}_{\text{exc}}(0, z_0) \;.
\ee
where, $\varepsilon_0$ is the electric permittivity of free space, $\alpha_0$ is the polarizability of the spherical particle given by: $\alpha_0  = 4 \pi a^3(\varepsilon_p - \varepsilon_1)/(\varepsilon_p + 2 \varepsilon_1)$;   $\varepsilon_p$ is the dielectric permittivity of the particle, and $\mathbf{E}_{\text{exc}}$ is the exciting field \textit{external} to the particle and comprises of not just the incident field $\mathbf{E}_i$ but also the radiative reaction field (scattered field)\cite{draine1988discrete} from the particle. If we use the (dyadic) Green function which relates an electric-dipole source $\mathbf{p}(\mathbf{r'})$ at a position $\mathbf{r'}$ to the electric field $\mathbf{E}(\mathbf{r})$ at a position $\mathbf{r}$ through the relation $\mathbf{E}(\mathbf{r}) = \tnsr{\mathbf{G}}(\mathbf{r},\mathbf{r'}) \mathbf{p}(\mathbf{r'})$ we can write the exciting field $\mathbf{E}_{\text{exc}}$ as:
\be
\label{Eexc}
\mathbf{E}_{\text{exc}}(\mathbf{r}) = \mathbf{E}_{\text{i}}(\mathbf{r}) + \tnsr{\mathbf{G}}_0(\mathbf{r},\mathbf{r'}) \mathbf{p}_0(\mathbf{r'})
\ee
where in the limit of $a (\omega/c) \rightarrow 0$ the Green's function $\tnsr{\mathbf{G}}_0(\mathbf{r},\mathbf{r'})$ is given by \cite{carminati2006radiative}:
\be
\label{G0}
\tnsr{\mathbf{G}}_0(\mathbf{r},\mathbf{r'}) \approx i \frac{k_1^3}{6 \pi \varepsilon_0} \tnsr{\mathbf{I}}  - \frac{1}{3 \varepsilon_0 \varepsilon_1} \delta(\mathbf{r} - \mathbf{r'})\tnsr{\mathbf{I}}
\ee
with $\tnsr{\mathbf{I}}$ being the unit dyad. If we express the dipole moment $\mathbf{p}_0$ in terms of the incident field $\mathbf{E}_{\text{i}}$ as:
\be
\label{p01}
\mathbf{p}_0 = \varepsilon_0 \alpha_{\text{eff}} \mathbf{E}_{i}(0, z_0) \;.
\ee
from Eq. \ref{p0}, \ref{Eexc} and \ref{G0} we get:
\be
\label{alphaEff}
\alpha_{\text{eff}} = \frac{\alpha_0}{1 - i k_1^3/(6 \pi)\alpha_0 }
\ee

It must be noted that the expression for $\alpha_{\text{eff}}$ in Eq. \ref{alphaEff} is valid in the limit $a\, \omega/c \rightarrow 0$. For larger particles an expression for $\alpha_{\text{eff}}$ can be derived from Mie theory of scattering from a sphere and is given by \cite{langlais2014cooperative, jylha2006modeling}:
\be
\label{alphaEff1}
\alpha_{\text{eff}} = \frac{6 \pi}{k_0^3 \sqrt{\varepsilon_1} (C_E - i) }
\ee
where the coefficient $C_E$ is:
\be
C_E = \frac{\left(\dfrac{g_m^2 - g_h^2}{g_m^2 g_h^2} \right)(\cos g_h + g_h \sin g_h) (\sin g_m - g_m \cos g_m) + g_m \cos g_h \cos g_m + g_m \sin g_h \sin g_m}{\left(\dfrac{g_h^2 - g_m^2}{g_m^2 g_h^2} \right) (\sin g_h - g_h \cos g_h) (\sin g_m - g_m \cos g_m) - g_m \sin g_h \cos g_m + g_h \cos g_h \sin g_m}
\ee
with $g_h$ and $g_m$ being non-dimensional factors given by  $\sqrt{\epsilon_1}(\omega/c) a  $ and $ \sqrt{\epsilon_p}(\omega/c) a $ respectively.
The scattered field from the polarized particle with dipole moment $\mathbf{p}_0 $ given from Eq. \ref{p01} can be written in Sipe's formalism \cite{sipe1987new} as:
\be
\label{sca_field}
\mathbf{E} (\mathbf{r}, z)   =    \int_{\mathbf{k_{\parallel}}} \frac{d^2\mathbf{k_{\parallel}}}{(2 \pi)^2}  \mathbf{F} (\mathbf{k_{\parallel}}, z-z_0) e^{i \mathbf{k_{\parallel}}. \mathbf{r}}
\ee
where,
\begin{align}
\label{Fs12}
\mathbf{F}(\mathbf{k_\parallel} ; z - z_0 > 0) &= \frac{i}{2} \left(\frac{\varepsilon_1}{\varepsilon_0} \right)  \left( \frac{\omega}{c} \right)^2 \frac{e^{i{k}_{z1} (z-z_0)}}{k_{z1}}
 \Big[  \uv{s}\uv{s} +  \uv{p}_{1+}\uv{p}_{1+}  \Big] \cdot \mathbf{p}_0  \\
 \mathbf{F}(\mathbf{k_\parallel} ; z - z_0 < 0) &= \frac{i}{2} \left(\frac{\varepsilon_1}{\varepsilon_0} \right)  \left( \frac{\omega}{c} \right)^2 \frac{e^{i{k}_{z1} (z-z_0)}}{k_{z1}}
 \Big[  \uv{s}\uv{s} +  \uv{p}_{1-}\uv{p}_{1-}  \Big] \cdot \mathbf{p}_0  \\
\end{align}

The key point here is that this form of the scattered field satisfies energy conservation and reciprocity. We now proceed to derive a model for interface scattering using this result as a building block.

\section{\label{sec:eff_index} Effective index model}
For a layer of particles located in a homogeneous medium it is possible to derive the specular reflection and transmission coefficients of an incident planar wave by finding the mean scattered field from the particles (as shown in \ref{sec:microscopic}). However, while this derivation throws light on the coherent effects in scattering by the particles, it is not convenient to adopt this approach for finding the coefficients when interfaces are present in the surrounding medium. To account for multiple reflections between the interfaces and the layer of particles, we develop an alternative approach based on the effective index theory to derive the reflection and transmission coefficients for the case of a layer of particles in a homogeneous medium,  show that the expressions for the coefficients are equivalent to that derived from the mean-field theory, and look for ways to extend it to the case where interfaces are present.

In this method we use the Maxwell Garnet theory to replace the layer of particles by a thin film of arbitrarily small thickness and with a refractive index such that the total polarization in the film is the same as that of the layer of particles. A plane wave would thus interact with such a film in the same way as it would with the layer of particles. If $\varepsilon_{\text{eff},x}$, $\varepsilon_{\text{eff},y}$ and $\varepsilon_{\text{eff},z}$ are the dielectric properties of the film along the $x$, $y$ and $z$ directions respectively,   $E_x$, $E_y$ and $E_z$ are the components of the electric field above the effective index layer in the incident plane (here the boundary conditions necessitates that $E_x$ and $E_y$ are the same in the film and outside), $\rho$ is the density of particles (units of $\text{m}^{-2}$) in the layer and $d_{\text{eff}}$ is the thickness of the film, equating the total polarization per unit area of the film to the total polarization in the layer of the particles we get:
\begin{align*}
 (\varepsilon_{\text{eff},y} - 1)E_y d_{\text{eff}} &=\rho \alpha_{\text{eff}} E_y \tab \text{for} \;\; \text{TE} - \text{polarization}  \\
 (\varepsilon_{\text{eff},x} - 1)E_x d_{\text{eff}}    &=\rho \alpha_{\text{eff}} E_x \tab \text{for} \;\; \text{TM} - \text{polarization} \\
 ( \varepsilon_{\text{eff},z} - 1)\dfrac{E_z \varepsilon_{1}}{\varepsilon_{\text{eff},z}} d_{\text{eff}} &=  \rho \alpha_{\text{eff}} E_z \tab \text{for} \;\; \text{TM} - \text{polarization}
\end{align*}
This gives us the dielectric properties of the film as:
\begin{align}
\label{dielectric_an}
\varepsilon_{\text{eff},x} = \varepsilon_{\text{eff},y} &= (1 + \rho \alpha_{\text{eff}}/ d_{\text{eff}}) \nonumber \\
 \varepsilon_{\text{eff},z} &= 1/\left( 1 - \rho \alpha_{\text{eff}}/(\varepsilon_1 d_{\text{eff}})\right).
 \end{align}
  The polarizability of the particles, $\alpha_{\text{eff}}$, is isotropic for the case of particles located in a homogeneous medium (this condition, however, gets relaxed when interfaces are present as discussed in Sec. \ref{sec:layer_substrate}). The reflection and transmission coefficients of a plane wave incident on the thin film is given by:
\begin{equation}
\label{RthinFilm}
r_{\text{eff}} = \frac{ r -  r \,  e^{2 i k_{\text{eff}} d_{\text{eff}}}}{1  - r^2 \,  e^{2 i k_{\text{eff}} d_{\text{eff}}} }
\end{equation}
and
\begin{equation}
\label{TthinFilm}
t_{\text{eff}} = \frac{ t^{(1)}   t^{(2)}  e^{2 i k_{\text{eff}} d_{\text{eff}}}}{1  - r^2 \,  e^{2 i k_{\text{eff}} d_{\text{eff}}} }
\end{equation}
where $k_{\text{eff}} = \sqrt{\varepsilon_{\text{eff},y} (\omega/c)^2 - \mathbf{k}_{\text{inc}}^{|| 2 } } $, $r$ is the Fresnel reflection coefficient which for a TE mode is given by:
\be
r^{\text{TE}} = \frac{k_{z1} - k_{\text{eff}}}{k_{z1} + k_{\text{eff}} }
\ee
and $t^{(1)}$ and $t^{(2)}$ are the Fresnel transmission coefficients given by (for TE mode):
\be
t^{(1)} = \frac{2 k_{z1} }{k_{z1} + k_{\text{eff}} }
\ee
\be
t^{(2)} = \frac{2 k_{\text{eff}} }{k_{z1} + k_{\text{eff}} }
\ee
In the limit $d_{\text{eff}}/(\rho \alpha_0) \rightarrow 0$ making the approximations $k_{\text{eff}} \approx n_{\text{eff}}\,\omega/c$; $(e^{2 i k_{\text{eff}} d_{\text{eff}}} - 1)  \approx 2 i\, n_{\text{eff}} \,(\omega/c) d_{\text{eff}}$; and $n_{\text{eff}} \approx \sqrt{ \rho \alpha_{\text{eff}}/d_{\text{eff}}}$ in Eq. \ref{RthinFilm} we can show that:
\be
\label{R_TE_2}
r_{\text{eff}} \approx \frac{ \rho \dfrac{i}{2} \left( \dfrac{\omega}{c} \right)^2 \dfrac{\varepsilon_1 \alpha_{\text{eff}}}{k_{z1}} }{1 - \rho \dfrac{i}{2} \left( \dfrac{\omega}{c} \right)^2 \dfrac{\varepsilon_1 \alpha_{\text{eff}}}{k_{z1}} }
\ee
and
\be
\label{T_TE_2}
t_{\text{eff}} \approx \dfrac{1}{1 - \rho \dfrac{i}{2} \left( \dfrac{\omega}{c} \right)^2 \dfrac{\varepsilon_1 \alpha_{\text{eff}}}{k_{z1}} }
\ee
The corresponding coefficients for the TM mode are slightly more complicated on account of the different dielectric properties along the $x$ and $z$ directions, and hence the coefficients do not reduce to simple expressions like in Eq. \ref{R_TE_2} and \ref{T_TE_2}. The form of the Fresnel coefficients to be used in Eq. \ref{RthinFilm} and \ref{TthinFilm} for the TM modes are given in \ref{sec:anisotropy}. 
The expressions for the reflection and transmission coefficients shown in Eq. \ref{R_TE_2} - \ref{T_TE_2} are equivalent to the expressions for the coefficients that are derived by considering the mean scattered field of individual particles in the layer (see \ref{sec:microscopic})
%
%
  thereby implying that the \textit{coherent scattering} by a layer of particles can be equivalently modeled by a thin film of a fictitious material whose dielectric property is given by Eq. \ref{dielectric_an}.
   Thus from the energy conservation statement across a thin dielectric film  it is possible to arrive at an expression for energy conservation across the layer of particles, as follows. The energy conservation statement across a thin dielectric film reads:
 \be
  \label{energy_consvn0}
 \text{Re} ( \mathbf{S}_1 \cdot \hat{\mathbf{z}}) = \text{Re} ( \mathbf{S}_2 \cdot \hat{\mathbf{z}}) + \int \text{Re}\Big[- \frac{i \omega \varepsilon_0}{2} (\tnsr{\varepsilon_{\text{eff}}} -1)\mathbf{E}_d(\mathbf{r}, z). \mathbf{E}_d^*(\mathbf{r}, z) \Big] \, \, dz
  \ee
  where $\mathbf{S}_1$ and $\mathbf{S}_2$ are the Poynting vectors of the coherent field at interfaces above and below the layer of particles in the homogeneous medium respectively and the integral term in Eq. \ref{energy_consvn0} represents the extinction in the coherent field across the layer of the particles, with  $\mathbf{E}_d(\mathbf{r}, z)$ being the field inside the thin-film, $\tnsr{\varepsilon_{\text{eff}}}$ is a diagonal tensor with elements $(\varepsilon_{\text{eff}, x}, \varepsilon_{\text{eff}, y}, \varepsilon_{\text{eff},z})$;  `*' indicates the complex conjugate and $dz$ being a unit element of the effective index layer along the $z$ axis. The integral is over the thickness of the thin dielectric film. Eq. \ref{energy_consvn0} can be shown to reduce to the form:
 \be
\label{energy_consvn}
\frac{k_{1z} }{2\omega \mu_0} (|r_{\text{eff}}|^2+ |t_{\text{eff}}|^2 - 1 ) =   \int \text{Re}\Big[- \frac{i \omega \varepsilon_0}{2} (\tnsr{\varepsilon_{\text{eff}}} -1)\mathbf{E}_d(\mathbf{r}, z). \mathbf{E}_d^*(\mathbf{r}, z) \Big] \, \, dz
 \ee
where, $r_{\text{eff}}$ and $t_{\text{eff}}$ are the reflection and transmission coefficients from Eq. \ref{R_TE_2} and \ref{T_TE_2}. Substituting the values from Eq. \ref{dielectric_an} and taking  the limit $d_{\text{eff}}/(\rho \alpha_0) \rightarrow 0$ we can write:

\begin{equation}
\label{RHS}
 \int \text{Re}\Big[  - \frac{i \omega \varepsilon_0}{2} (\tnsr{\varepsilon_{\text{eff}}} -1)\mathbf{E}_d(\mathbf{r}, z_0). \mathbf{E}_d^*(\mathbf{r}, z_0)\Big] \, \, dz = \text{Re} \Big[ - \frac{i \omega \varepsilon_0}{2}  \rho \alpha_{\text{eff}}  |\mathbf{E}_{\text{ill}}(\mathbf{r}, z_0)|^2 \Big]
\end{equation}
where $\mathbf{E}_{\text{ill}}(\mathbf{r}, z_0)$ is the field outside the film thus comprising of both the incident and reflected fields.

The right hand side of Eq. \ref{energy_consvn} is the extinction in the coherent component of the incident radiation, which comprises of both the diffusely scattered radiation as well as the flux absorbed by the particles in the layer. This can be seen by comparing the expressions for the power scattered and absorbed by a single particle.  The power absorbed by a particle in a homogeneous medium is given by \cite{carminati2006radiative}:
\be
\label{power_abs_particle}
W_{\text{abs}} = \frac{\omega \epsilon_0}{2} \left( \text{Im}(  \alpha_{\text{eff}}) - \frac{k_1^3}{6 \pi}|\alpha_{\text{eff}}|^2  \right) |\mathbf{E}_\text{ill}|^2;
\ee
The field scattered by a dipole in a homogeneous medium is given by Eq. \ref{sca_field} from which we can show the power scattered to be \cite{bohren98a}:
\be
\label{power_sca_particle}
W_{\text{sca}} = \frac{\omega \epsilon_0}{2} \left( \frac{k_1^3}{6 \pi} | \alpha_{\text{eff}}|^2 \right) |\mathbf{E}_\text{ill}|^2;
\ee
From Eq. \ref{energy_consvn},  \ref{RHS}, \ref{power_abs_particle} and \ref{power_sca_particle} we get the energy conservation statement across the layer of particles as:
\be
\label{extinction}
\frac{k_{1z} }{2\omega \mu_0} (|r_{\text{eff}}|^2+ |t_{\text{eff}}|^2 - 1 ) = \rho W_{\text{sca}} + \rho  W_{\text{abs}}
\ee
with $\rho W_{\text{sca}}$ accounting for the total diffuse scattered power from the particles. It is worth pointing out explicitly that the illuminating field  $\mathbf{E}_\text{ill}$ includes not just the incident field $\mathbf{E}_\text{i}$ from Eq. \ref{Inc} but also the coherently scattered field from the particles. This can also be seen in the microscopic description of scattering from the layer of particles shown in \ref{sec:microscopic} where the mean scattered field from the particles has been shown to account for the coherently reflected field.   The dipole moment of the particles in the layer, $\mathbf{p}_0$, can thus be written as:
\be
\label{p0_layer}
\mathbf{p}_0 = \varepsilon_0 \alpha_0 (\mathbf{E}_{\text{exc}} + \mathbf{E}_{\text{refl} }) \;,
\ee
where $\mathbf{E}_{\text{refl}}$ is the coherently reflected field from the layer of particles and $\mathbf{E}_{\text{exc}}$ is given from Eq. \ref{Eexc}. As such, while the diffuse scattered power $\rho W_{\text{sca}} $ in Eq. \ref{extinction} seems to resemble the independent-scattering approximation \cite{durant2007light}, we point out that the model does take into account the interaction between the particles in the mean-field approximation. However the model does not take into account the near-field interactions and correlations in the position of the particles. The model inherently satisfies reciprocity principle since it makes use of an analytic solution of the electromagnetic scattered field from a particle which is known to satisfy reciprocity \cite{mishchenko2002scattering}. This will be shown explicitly in Sec.  \ref{sec:LayerOfparticlesAboveSub}.

\section{ \label{sec:layer_substrate}Analysis of coherent scattering from a mono-layer of particles above a substrate}
\begin{figure}[Hhbt]
\begin{center}
\includegraphics[trim = 0cm 0cm 0cm 0cm, clip=true, scale=0.4]{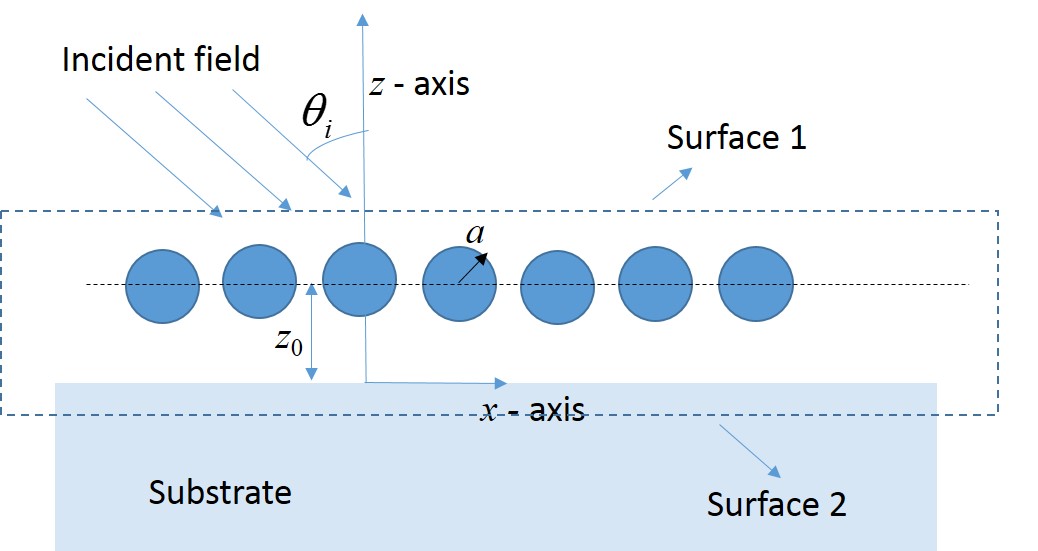}
\includegraphics[trim = 0cm 0cm 0cm 0cm, clip=true, scale=0.4]{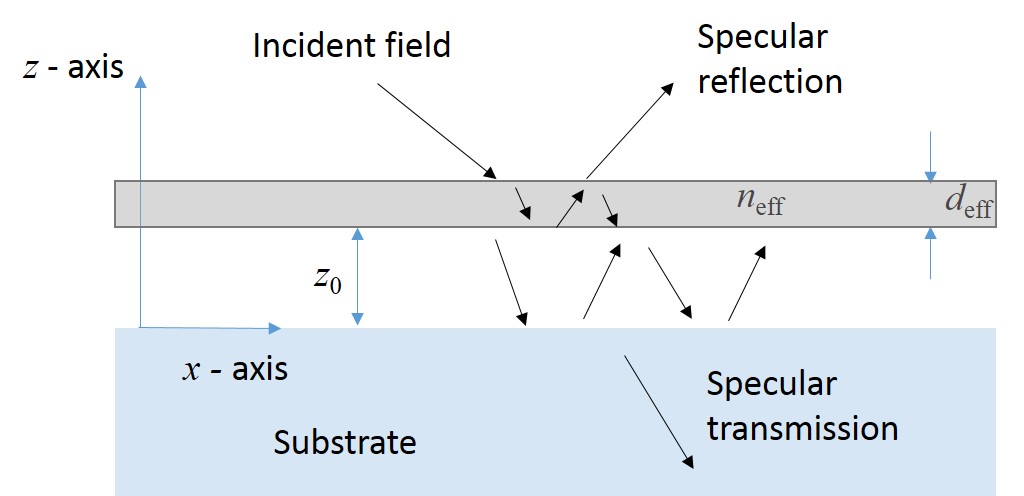}
\caption{(a) Configuration for finding the scattered field from a layer of particles located at a distance $z_0$ in a homogeneous medium of dielectric function $\epsilon_1$ on top of a substrate with dielectric function $\epsilon_2$. Surfaces 1 and 2 are located just above the position of the particles at $z=z_0^+$ and just below the surface of the substrate at $z=0^-$ respectively. (b) In order to take into account multiple reflections between the layer of particles and the substrate the layer of particles is modeled as a thin film of arbitrarily small thickness $d_{\text{eff}}$ with effective dielectric function $\varepsilon_{\text{eff}}$ and located at a distance $z_0$ from the substrate }
\label{layer_substrate}
\end{center}
\end{figure}

As explained in Sec. \ref{intro} the intention of this work is to develop a BSDF of a surface which satisfies energy conservation and reciprocity. With this aim in mind and with the results from the previous sections we now attempt to find the BSDF of a surface with a layer of particles on top of it acting as the source of scattering as shown in Fig. \ref{layer_substrate}(a).
Scattering from a single particle on top of a substrate has been discussed in detail in Ref. \cite{dahan2012enhanced}. 
The main results from this analysis is presented in Sec. \ref{sec:particleAboveSub} and the results are utilized to extend the analysis to the case of a layer of particles on top of a substrate in Sec.  \ref{sec:LayerOfparticlesAboveSub}.

\subsection{ \label{sec:particleAboveSub} A single particle above a substrate}
Consider the case of a single particle located at $(0, z_0)$ above a substrate comprising the half-space $z<0$.
In the vicinity of the substrate the radiation reaction field from the particle discussed in Sec. \ref{sec:single} will be altered due to reflection of the scattered field from the substrate below. The effective polarizability  from Eq. \ref{alphaEff}  will thus have to be modified to take into account this reflected component. The procedure to arrive at this modified form of $\alpha_{\text{eff}}$ is similar to the one detailed in Sec. \ref{sec:single} with one difference - the effective polarizability will now be a tensor. The polarizability tensor $\tnsr{\alpha}_{\text{eff}}$ is diagonal with elements $(\alpha^{\text{xx}}_{\text{eff}}, \alpha^{\text{yy}}_{\text{eff}}, \alpha^{\text{zz}}_{\text{eff}})$ and with $\alpha^{\text{xx}}_{\text{eff}} = \alpha^{\text{yy}}_{\text{eff}} \neq \alpha^{\text{zz}}_{\text{eff}}$ as the presence of the surface breaks the rotational symmetry.   The exciting field $\mathbf{E}_{\text{exc}}$  in Eq. \ref{Eexc} is now given by:
\be
\label{Eexc2}
\mathbf{E}_{\text{exc}}(\mathbf{r}) = \mathbf{E}_i(\mathbf{r}) + [\tnsr{\mathbf{G}}_0(\mathbf{r},\mathbf{r}_0) + \tnsr{\mathbf{G}}_r(\mathbf{r},\mathbf{r}_0)] \mathbf{p}_0(\mathbf{r}_0)
\ee
where $\tnsr{\mathbf{G}}_r$ is the (dyadic) Green's function which accounts for the reflection component of the scattered field from the interface and can be calculated by the Fourier transform \cite{dahan2012enhanced}:
\be
\label{G_rr}
\tnsr{\mathbf{G}}_r(\mathbf{r}, \mathbf{r}_0) = \int_{-\infty}^{+\infty} \frac{ d^2\mathbf{k}^{|| }}{(2 \pi)^2}\tnsr{\mathbf{G}}_r(\mathbf{k}^{|| }; z - z_0)e^{i \mathbf{k}^{|| }. \mathbf{R}}
\ee
with
\be
\label{G_rk}
\tnsr{\mathbf{G}}_r(\mathbf{k}^{|| }; z - z_0)  = \frac{i}{2} \left(\frac{\varepsilon_1}{\varepsilon_0} \right)  \left(\frac{\omega}{c} \right)^2 \frac{e^{i k_{z1}  (z+z_0)}}{k_{z1} }\Big[ r_{12}^{\text{TE}}\, \uv{s} \uv{s}  +  r_{12}^{\text{TM}} \uv{p}_{1+}  \uv{p}_{1-}  \Big]
\ee
Here, $r_{12}^{\text{TE}}$ and $r_{12}^{\text{TM}}$ are the Fresnel reflection coefficients for the TE and TM modes respectively at the interface between the homogeneous medium of dielectric function $\varepsilon_1$ and the substrate. The dipole moment of the particle is given by:
\be
\label{p02}
\mathbf{p}_0 = \varepsilon_0 \tnsr{\boldsymbol{\alpha}}_{\text{eff}}\mathbf{E}_{i}(0, z_0) \;.
\ee
and from Eqs.  \ref{p0}, \ref{G0}, \ref{Eexc2}, and \ref{p02} it is possible to show that the effective polarizability $\tnsr{\boldsymbol{\alpha}}_{\text{eff}}$ takes the form:
\be
\label{alphaEff2}
\tnsr{\boldsymbol{\alpha}}_{\text{eff}} = \alpha_0 \Bigg[\tnsr{\boldsymbol{\text{I}}} - \left(i k_1^3/(6 \pi)\tnsr{\boldsymbol{\text{I}}} + \tnsr{\mathbf{G}}_r \right) \alpha_0 \Bigg]^{-1}
\ee
where $\tnsr{\boldsymbol{\text{I}}}$ is the unit dyad.

The scattered power $W_\text{sca}$ from a particle with dipole moment $\mathbf{p}_0$ given in Eq. \ref{p02} is no longer independent of the orientation of the dipole. If $W_\text{sca}^{(\text{R})}$ denotes the scattered power from the particle at a surface just above the position of the particle at at $z= 0^+$,  and if $W_\text{sca}^{(\text{T})}$ denotes the scattered power from the particle at a surface just below the surface of the substrate at $z= 0^-$,   
 expressions to calculate $W_\text{sca}^{(\text{R})}$ and $W_\text{sca}^{(\text{T})}$ as an integral over $\mathbf{k}_{||}$  have been derived in Ref. \cite{dahan2012enhanced}. These are given by:

\be
 \begin{split}
 \label{W_sca_1}
W_{sca}^{(\text{R})}(z = z_0^+) = - \frac{1}{2 \omega \mu_0}  \text{Re} \Bigg[    \int_{\mathbf{k}_{||}}  \frac{d^2 \mathbf{k}_{||} }{(2 \pi)^2} \mathbf{F}_{si}(\mathbf{k}_{||};z_0)   \times \left(  \mathbf{k}_{i}^* \times   \mathbf{F}_{si}^*(\mathbf{k}_{||};z_0)  \right) \Bigg]. \hat{\mathbf{z}}
\end{split}
\ee

\be
 \begin{split}
 \label{W_sca_2}
W_{sca}^{(\text{T})}(z = 0^-) = \frac{1}{2 \omega \mu_0}  \text{Re} \Bigg[   \int_{\mathbf{k}_{||}}  \frac{d^2 \mathbf{k}_{||} }{(2 \pi)^2} \mathbf{F}_{st}(\mathbf{k}_{||};z_0)   \times \left(  \mathbf{k}_{t}^* \times   \mathbf{F}_{st}^*(\mathbf{k}_{||};z_0)   \right) \Bigg] . \hat{\mathbf{z}}
\end{split}
\ee
with
\be
\label{Fsi}
\mathbf{F}_{si}(\mathbf{k}_{||};z_0)  = \frac{i}{2} \left(\frac{\varepsilon_1}{\varepsilon_0} \right)  \left(\frac{\omega}{c} \right)^2 \frac{1}{k_{z1}}  \Big[ \left( 1 + r_{12}^{\text{TE}}  e^{2 i k_{z1} z_0}  \right) \, \uv{s} \uv{s}  +  \left( \uv{p}_{1+}  \uv{p}_{1+}   +   r_{12}^{\text{TM}}  e^{2 i k_{z1} z_0}  \uv{p}_{1+}  \uv{p}_{1-}   \right) \Big] \mathbf{p}_0
\ee
and
\be
\mathbf{F}_{st}(\mathbf{k}_{||};z_0)  = \frac{i}{2} \left(\frac{\varepsilon_1}{\varepsilon_0} \right)  \left(\frac{\omega}{c} \right)^2 \frac{e^{i k_{z1}  z_0}}{k_{z1} }\Big[ t_{12}^{\text{TE}}\, \uv{s} \uv{s}  +  t_{12}^{\text{TM}}  \uv{p}_{2-}  \uv{p}_{1-}  \Big] \mathbf{p}_0
\ee
Here $t_{12}^{\text{TE}}$ and $t_{12}^{\text{TM}}$ are the Fresnel transmission coefficients for the TE and TM modes at the interface between the homogeneous medium of dielectric function $\varepsilon_1$ and the substrate respectively. The total power scattered from the particle $W_{\text{sca}}$ can be computed $W_\text{sca} = W_\text{sca}^{(\text{R})} + W_\text{sca}^{(\text{T})} $.  Note that Eq. \ref{Fsi} includes interference between the light scattered upward from the particle and the light scattered downward from the particle and subsequently reflected from the substrate below.  Another important point to be noted here is that the proximity of the particle to the substrate induces contributions from large wave-vectors ($k_{||} > \sqrt{\varepsilon_2} \, \, \omega/c$) to the scattering power $W_\text{sca}^{(\text{T})}$ which will be seen as absorptive losses in the substrate \cite{dahan2012enhanced}.  


\subsection{ \label{sec:LayerOfparticlesAboveSub} Layer of particles above a substrate}


With the results from Sec. \ref{sec:eff_index} and Sec. \ref{sec:particleAboveSub} we are now in a position to describe scattering from a layer of particles above a substrate. To obtain the coherent reflection and transmission coefficients for the configuration shown in Fig. \ref{layer_substrate}(a) of a plane wave incident on a substrate with particles on top of it, we adopt the procedure detailed in Sec. \ref{sec:eff_index} and replace the layer of particles with a thin film of thickness $d_\text{eff}$ as shown in Fig. \ref{layer_substrate}(b). To make sure that the total polarization is the same in both the cases and noting the anisotropic nature of the polarizability of the particles, the dielectric properties of the thin-film are now given by: 
\begin{align}
\label{dielectric_an1}
\varepsilon_{\text{eff},x} = \varepsilon_{\text{eff},y} &= (1 + \rho \alpha^{\text{xx}}_{\text{eff}}/ d_{\text{eff}}) \nonumber \\
 \varepsilon_{\text{eff},z} &= 1/\left( 1 - \rho \alpha^{\text{zz}}_{\text{eff}}/(\varepsilon_1 d_{\text{eff}})\right).
 \end{align}
 With these changes and employing standard recursion relations \cite{chew95a} or transfer-matrix method \cite{born1999principles} we can compute the coherent reflection and transmission coefficients for the multi-layered configuration shown in Fig. \ref{layer_substrate}(b).  The energy conservation statement from Eq. \ref{extinction} is still valid for this configuration, with the dipole moment of the particles $\mathbf{p}_0$ to compute $W_\text{sca}$ being given by  $\mathbf{p}_0 = \varepsilon_0 \tnsr{\boldmath{\alpha_{\text{eff}}}} \mathbf{E}_{\text{ill}}$ where the illuminating field  $\mathbf{E}_{\text{ill}}$ is taken to be the field above the thin-film in Fig. \ref{layer_substrate}(b). This field accounts for the multiple reflections between the particles in the layer as well as with the substrate below.

%


In radiometry the BSDF is frequently used to characterize the scatter of optical radiation from a surface as a function of the angle of the incident and the scattered beam. It is given by the ratio of the scattered radiance to the incident irradiance and has units of $\text{sr}^{-1}$. The terms bi-directional reflectance distribution function (BRDF) and bi-directional transmittance distribution function (BTDF) are used when referred specifically to the reflected and transmitted scatter respectively. If $I^+(\mu, \phi)$ and  $I^-(\mu, \phi)$ are the specific intensities of the scattered beam (units of W$\text{m}^{-2}\text{sr}^{-1}$) at interfaces 1 and 2 respectively shown in Fig.  \ref{layer_substrate}(a) where $\phi$ is the azimuthal angle and  $\mu = \cos \theta$ with $\theta$ representing the polar angle of the scattered beam, we can express the BRDF and BTDF for the surface shown in Fig. \ref{layer_substrate}(a) on which a coherent beam given in Eq. \ref{Inc} is incident upon it as:
\begin{align}
\label{BRDF}
\text{BRDF}(\mu, \phi) = \frac{(2 \omega \mu_0) \, I^+(\mu, \phi)}{\text{Re}\big[(\mathbf{E}_i  \times (  \mathbf{k}_{i}^* \times \mathbf{E}_i^* )) .  \hat{\mathbf{z}}\big]} \nonumber \\
\text{BTDF}(\mu, \phi) = \frac{(2 \omega \mu_0) \,  I^-(\mu, \phi)}{ \text{Re}\big[(\mathbf{E}_i  \times (  \mathbf{k}_{i}^* \times \mathbf{E}_i^* )) .  \hat{\mathbf{z}}\big]}
\end{align}
 To arrive at an expression for $I^+(\mu, \phi)$ and $I^-(\mu, \phi)$  in Eq. \ref{BRDF} in terms of the electromagnetic description given in the previous sections we compare the expressions for the reflected flux above the surface. If $\text{Flux}^{(\text{R})}$ denotes the radiative flux at surface 1 in Fig. \ref{layer_substrate}(a)  it is given in terms of $I^+(\mu, \phi)$ as:
\be
\label{flux_rt}
\text{Flux}^{(\text{R})} =  \int_{\phi = 0}^{2 \pi} \int_{\mu=0}^{1} I^+(\mu, \phi) \mu \,  d \mu \, d \phi
\ee

But from Eq. \ref{extinction} and the description given in Sec. \ref{sec:particleAboveSub} 
this must be equivalent to the diffuse scattering from the layer of particles at surface 1 which is given by:
\be
\label{flux_em}
\text{Flux}^{(\text{R})} = \rho W_{\text{sca}}^{(\text{R})}
\ee
with the integral over $k_{||}$ in Eq. \ref{W_sca_1} extending from 0 to $ \sqrt{\varepsilon_1} (\omega/c)$.
Utilizing the cylindrical transformation $d^2 \mathbf{k}_{||} = k_{||}\,d k_{||}\, d\phi$ in Eq. \ref{W_sca_1} and noting that  $k_{||} = k_1 \sin \theta$ we get $d^2 \mathbf{k}_{||} = k_1^2 \mu \, d \mu \, d \phi$. Thus Eq. \ref{W_sca_1} can be written as:
\be
 \begin{split}
 \label{W_sca_1_rt}
W_{\text{sca}}^{(\text{R})}(z = z_0^+) = \int_0^{2 \pi}  \int_0^1 M^+(\mu, \phi) k_1^2 \mu \, d \mu \, d \phi
\end{split}
\ee
with
\be
 M^+(\mu, \phi) = - \frac{1}{2 \omega \mu_0}  \text{Re} \Bigg[      \frac{1 }{(2 \pi)^2} \Big[ \mathbf{F}_{si}(\mathbf{k}_{||};z_0)   \times \left(  \mathbf{k}_{i}^* \times   \mathbf{F}_{si}^*(\mathbf{k}_{||};z_0)  \right) \Big] \Bigg]. \hat{\mathbf{z}}
\ee
From Eqs. \ref{flux_rt}, \ref{flux_em} and \ref{W_sca_1_rt}, we get:
\be
\label{Int_em}
I^+(\mu, \phi) = \rho M^+(\mu, \phi) k_1^2
\ee
Substituting this in Eq. \ref{BRDF} we get for the BRDF:
\begin{equation}
\label{BRDF1}
\text{BRDF}(\mu, \phi) = \frac{(2 \omega \mu_0) \, \rho M^+(\mu, \phi) k_1^2}{\text{Re}\big[(\mathbf{E}_i  \times (  \mathbf{k}_{i}^* \times \mathbf{E}_i^* )) .  \hat{\mathbf{z}}\big]} \\
\end{equation}
Similarly, the expression for BTDF reads:
\begin{equation}
\label{BTDF1}
\text{BTDF}(\mu, \phi) = \frac{(2 \omega \mu_0) \,\rho M^-(\mu, \phi) k_1^2}{\text{Re}\big[(\mathbf{E}_i  \times (  \mathbf{k}_{i}^* \times \mathbf{E}_i^* )) .  \hat{\mathbf{z}}\big]} \\
\end{equation}
where $M^-(\mu, \phi)$ is given by:
\be
 M^-(\mu, \phi) =  \frac{1}{2 \omega \mu_0}  \text{Re} \Bigg[      \frac{1 }{(2 \pi)^2} \Big[ \mathbf{F}_{st}(\mathbf{k}_{||};z_0)   \times \left(  \mathbf{k}_{t}^* \times   \mathbf{F}_{st}^*(\mathbf{k}_{||};z_0)  \right) \Big] \Bigg]. \hat{\mathbf{z}}
\ee



 The reciprocity relation for this model can be confirmed by comparing the differential scattering cross section in the scattering direction ($\theta_{\text{sc}}$,$\phi_{\text{sc}}$) for an incident beam at angle ($\theta_i$, $\phi_i$) in the configuration shown Fig. \ref{layer_substrate}, with that obtained by interchanging the incident and scattering directions. That is, for reciprocity to hold true we need \cite{schmidt2012case}:
\begin{equation}
\frac{d \sigma}{d \Omega}(\theta_i, \phi_i;  \theta_{\text{sc}}, \phi_{\text{sc}}) = \frac{d \sigma}{d \Omega}(\theta_{\text{sc}}, \phi_{\text{sc}};  \theta_{\text{i}}, \phi_i)
\label{DSCS}
\end{equation}
where $\dfrac{d \sigma}{d \Omega}$ represents the differential scattering cross section. In terms of the BRDF, Eq. \ref{DSCS} is equivalent to:
\begin{equation}
\text{BRDF}_\text{i}(\mu_\text{sc}, \phi_\text{sc})\, \mu_\text{sc} = \text{BRDF}_\text{sc}(\mu_i, \phi_i)\, \mu_i
\end{equation}
where $\text{BRDF}_\text{i}(\mu_\text{sc}, \phi_\text{sc})$ [$\text{BRDF}_\text{sc}(\mu_i, \phi_i)$] represents the BRDF from Eq. \ref{BRDF1} for light incident in the direction ($\theta_i$, $\phi_i$) [($\theta_\text{sc}$, $\phi_\text{sc}$)], $\mu_\text{sc} = \cos \theta_\text{sc}$, and  $\mu_i = \cos \theta_i$. A similar expression can also be written for the BTDF.
We confirm the reciprocity relation for our model by taking two pairs of arbitrary angles for the incident and scattering directions as shown in Table \ref{err:reciprocity} and finding the error calculated as $\Big[\dfrac{d \sigma}{d \Omega}(\theta_i, \phi;  \theta_{\text{sc}}, \phi) -\dfrac{d \sigma}{d \Omega}(\theta_{\text{sc}}, \phi;  \theta_{\text{i}}, \phi)\Big]/\Big[\dfrac{d \sigma}{d \Omega}(\theta_i, \phi;  \theta_{\text{sc}}, \phi) + \dfrac{d \sigma}{d \Omega}(\theta_{\text{sc}}, \phi;  \theta_{\text{i}}, \phi)\Big] $ for a fixed azimuthal angle $\phi = 0$. The  values shown are for the average of TE and TM polarizations. The material parameters in the configuration are arbitrarily chosen  with the polarizability of the particles $\alpha_0 = 6.06 \times 10^6 \,\, \text{nm}^3$, dielectric function of the substrate $\varepsilon_2 = 3.91+1.2i$, incident beam of wavelength $\lambda = 300$ nm and filling fraction of the particles $f = 0.15$.  The non-dimensional quantity $f$ is related to the density of the particles $\rho$ as $f = \rho \pi a^2$. The reciprocity relation is observed to be satisfied up to floating point precision.
\begin{table}[h]
\centering
\begin{tabular}{|r|r|}
\hline
$\theta_i$, $\theta_{\text{sc}}$ & error  \\
\hline
 40$^\circ$, 74$^\circ$ & $10^{-14}$  \\
\hline
 58.3$^\circ$, 85.1$^\circ$& $10^{-14}$  \\
\hline
\end{tabular}
\caption{Error in satisfying the reciprocity relation in our  model for two pairs of arbitrarily chosen angles}
\label{err:reciprocity}
\end{table}

Noting that the integral over the azimuthal angle $\phi$ in Eq. \ref{flux_rt} can be analytically obtained, the scattered flux in Eq. \ref{flux_rt} can be alternatively expressed as:
\be
\label{flux_rt2}
\text{Flux}^{(\text{R})} = \int_\mu I_{\text{mod}}^+(\mu) \mu \, d \mu
\ee
where,  from Eq. \ref{flux_rt} and \ref{Int_em}, we have $I_{\text{mod}}^+(\mu) =  \int_0^{2 \pi}\rho M^+(\mu, \phi) k_1^2 d\phi $. We can use this to define the BRDF alternatively as:
\begin{equation}
\label{BRDF2}
\text{BRDF}(\mu) =  \frac{ (2 \omega \mu_0) \, \int_0^{2 \pi} \rho M^+(\mu, \phi) k_1^2 d\phi}{\text{Re}\big[(\mathbf{E}_i  \times (  \mathbf{k}_{i}^* \times \mathbf{E}_i^* )) .  \hat{\mathbf{z}}\big]} \\
\end{equation}
\begin{figure}[Hhbt]
\begin{center}
\includegraphics[trim = 0cm 0cm 0cm 0cm, clip=true, scale=0.45]{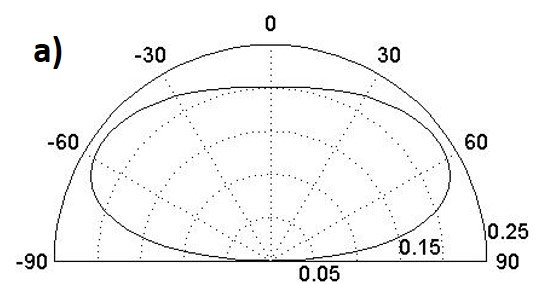}
\includegraphics[trim = 0cm 0cm 0cm 0cm, clip=true, scale=0.45]{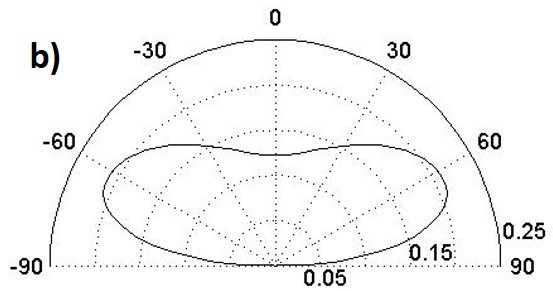}
\includegraphics[trim = 0cm 0cm 0cm 0cm, clip=true, scale=0.77]{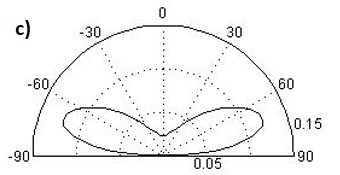}
\includegraphics[trim = 0cm 0cm 0cm 0cm, clip=true, scale=0.35]{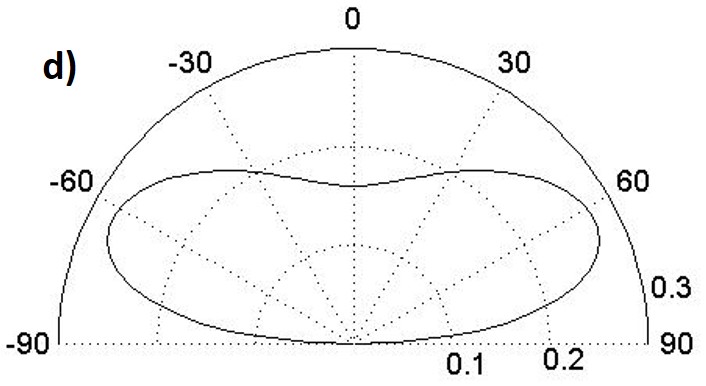}
\includegraphics[trim = 0cm 0cm 0cm 0cm, clip=true, scale=0.45]{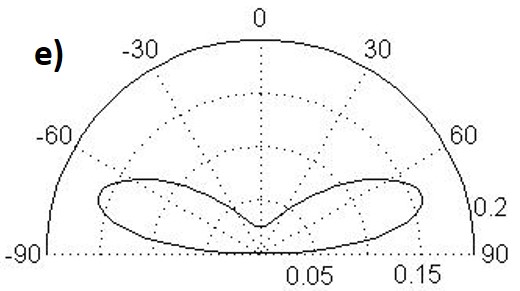}
\caption{ BRDF from Eq. \ref{BRDF2}  for the configuration shown in Fig. \ref{layer_substrate} with the following parameters:  $\varepsilon_2 = 3.91+1.2i$; $\lambda = 300$ nm;  $f = 0.15$;  $\alpha_0 = 6.06 \times 10^6 \,\, \text{nm}^3$ and (a) $z_0 = 85$ nm, (b)  $z_0 = 100$ nm,  and (c) $z_0 = 135$ nm. The parameters for Fig. (d) and (e) are the same as in (b) and (c) respectively except for $\alpha_0$ which is increased to $9.87 \times 10^6 \,\, \text{nm}^3$   }
\label{scatteringDF}
\end{center}
\end{figure}
We now have all the tools required to describe the utility of the scattering model developed so far. As a demonstration,  consider the particular case of scattering from the surface shown in  Fig. \ref{layer_substrate}(a) with the material properties taken to be the same as that chosen for showing reciprocity relation in Table \ref{err:reciprocity} and assuming normal incidence. The BRDF from Eq. \ref{BRDF2} is plotted in Fig.  \ref{scatteringDF}(a)-(c) as a function of the scattering angle $\theta$ for varying position of the particles above the substrate, $z_0$. What we observe is that by increasing $z_0$ scattering lobes with large angle scattering can be obtained. Likewise, keeping $z_0$ constant if we increase the polarizability  $\alpha_0$ of the particles (or equivalently the filling fraction $f$) we can increase the scattering cross section of the particles. This can be seen by comparing Fig. \ref{scatteringDF}(b) and \ref{scatteringDF}(d) (or similarly Fig. \ref{scatteringDF}(c) and \ref{scatteringDF}(e)) where $z_0$ has been kept constant but $\alpha_0$ has been increased. We observe that the magnitude of scattering has increased but the shape of the scattering profile has been retained. Thus by varying $z_0$ and $\alpha_0$ (or $f$) we can change both the scattering angle as well as magnitude of scattering and thus obtain a wide range of scattering profiles.  This can be used, for example, to analyze the effect of scattering in increasing the absorption of light in the active layer of a multilayered solar cell.  By tuning the parameters $z_0$ and $\alpha_0$ a wide range of scattering profiles can be explored so as to optimize absorption in the active layer.

\section{ \label{exact} Comparison with exact electromagnetic simulations}

The scattering model developed in this work can be compared with exact electromagnetic simulation of scattering by a random layer of spherical particles located on top of a finite substrate and averaged over many realizations. Details on the procedure for the exact electromagnetic simulations can be found in Ref. \cite{langlais2014cooperative}. It must be noted that the simulation will include the near-field interactions between the dipoles and hence is expected to diverge from the scattering model developed in this work for large filling fractions of the particles. In particular, it has been shown in an earlier work \cite{durant2007lightb} that near-field interactions and correlations in positions of the particles become important when the filling fraction of the particles is greater than approximately 5 \%.

Here we show a comparison between (i) specular reflection as predicted by the effective index model in Eq. \ref{RthinFilm}, and (ii) diffuse scattered power from Eq. \ref{flux_rt2} (both quantities normalized with the incident power) with the corresponding quantities observed from exact electromagnetic simulations.  This has been shown in Table \ref{tableComp} where we consider a particular case of scattering by a layer of particles lying on top of the substrate ($z_0 =a$) with arbitrarily chosen parameters: $\alpha_0 = 2.88 \times 10^6 \,\, \text{nm}^3$, $\varepsilon_2 =  3.91+1.2i$, $\lambda = 300$ nm,  and for different filling fractions 5\% and 15\%.   As expected the results deviate significantly for the higher filling fraction. The deviation for higher $f$ should not be viewed as a setback for our model since it was never our intention to model the multi-scattering problem accurately but rather we wished to formulate a scattering model which gives importance to satisfying energy conservation and reciprocity principles.
\begin{table}[h!]
\footnotesize
\begin{tabular}{|c|c|c|c|c|c|c|c|c|c|c|}
\hline
  \multirow{2}{*}{\textbf{a)}} & filling fraction = 5\% & \multicolumn{3}{|c|}{$\theta_i = 0^{\circ}$} & \multicolumn{3}{|c|}{$\theta_i = 30^{\circ}$}  & \multicolumn{3}{|c|}{$\theta_i = 60^{\circ}$} \\
\cline{3-11}
& $\alpha_0 = 2.88 \times 10^6 \,\, \text{nm}^3 $ & Exact& Approx& Error(\%)&  Exact& Approx& Error(\%)&   Exact& Approx& Error(\%) \\
\hline
 \multirow{3}{*}{TE} & SR without particles&0.1199 &0.1199 & -&0.1564  &0.1564 & -&0.3350 &0.3350&-\\
 & SR with particles & 0.0290  &  0.0313 & 7.9 & 0.0420 &0.0480 & 14.3 & 0.1600 & 0.1805& 12.5 \\
 & diffuse scattered power & 0.1160 & 0.1261& 8.6 & 0.1240& 0.1349 & 8.8 &0.1310 &0.1388&7.3 \\
\hline
 \multirow{3}{*}{TM} & SR without particles&0.1199 &0.1199 & - &0.0872 &0.0872 & - &0.0062&0.0062& -\\
 & SR with particles &0.0290 & 0.0313 & 7.9 & 0.0280 & 0.0317& 13.2 & 0.0120 &0.0115& 4.2 \\
 & diffuse scattered power &0.1160 & 0.1258 & 8.4 &0.1000& 0.1112 & 11.2&0.0890 & 0.1049&17.9 \\
\hline
\end{tabular}
\\
\\
\\
\\
\footnotesize
\begin{tabular}{|c|c|c|c|c|c|c|c|c|c|c|}
\hline
\multirow{2}{*}{ \textbf{b)}} & filling fraction = 15\% & \multicolumn{3}{|c|}{$ \theta_i = 0^{\circ}$} & \multicolumn{3}{|c|}{$\theta_i = 30^{\circ}$}  & \multicolumn{3}{|c|}{$\theta_i = 60^{\circ}$} \\
\cline{3-11}
& $\alpha_0 = 2.88 \times 10^6 \,\, \text{nm}^3$ & Exact& Approx& Error(\%)&  Exact& Approx& Error(\%)&   Exact& Approx& Error(\%) \\
\hline
 \multirow{3}{*}{TE} & SR without particles&0.1199 &0.1199 &- &0.1564 &0.1564 &- &0.3350&0.3350 &-\\
 & SR with particles & 0.0160 & 0.0094 & 41.3 & 0.0130 &0.0226& 73.8 & 0.0800 & 0.1533& 91.6\\
 & diffuse scattered power &0.1750  &0.2476 & 41.7&0.1890& 0.2577 &36.4& 0.2020& 0.2503 &23.9\\
\hline
 \multirow{3}{*}{TM} & SR without particles&0.1199 &0.1199 &- &0.0872& 0.0872&-& 0.0062& 0.0062&-\\
 & SR with particles &0.0148 &0.0094 & 36.5 &0.0025&0.0052& 108 &0.0200 &0.0249&24.5 \\
 & diffuse scattered power & 0.1800 &0.2477 &37.6 &0.1790& 0.2705 &50.2 &0.1900&0.3182&67.4\\
\hline
\end{tabular}
\caption{Normalized specularly reflected (SR) power from the effective index theory, and the diffuse scattered power from Eq. \ref{flux_rt2} (marked `Approx')  are compared with values from exact electromagnetic simulations (marked `Exact')  for the configuration shown in Fig. \ref{layer_substrate}(a) for different incident angles ($\theta_i$ = $0^{\circ}$, $30^{\circ}$ and $60^{\circ}$) and for two different filling fractions (a) 5 \% and (b) 15 \%. The error between the simulated and approximate values are expressed in terms of percentage of the exact values. Specularly reflected power from the substrate in the absence of the layer of particles is also presented for reference.}
\label{tableComp}
\end{table}


\section{Conclusion}

With the motive of exploring different scattering profiles to be used in optimizing absorption in the active layer of a multilayered solar cell, we have developed a scattering model based on scattering from a layer of particles on top of a substrate which satisfies both energy conservation and reciprocity and takes into account the particle-induced absorption in the substrate. In course of developing such a model we have shown that the effective index model based on Maxwell Garnet theory can sufficiently describe the coherent scattering from the layer of particles. While the intent of developing this model is not to accurately describe the scattering from a layer of particles, we have shown that in the regime where near-field interactions and correlation in the position of the particles are negligible the model sufficiently approximates both the coherent scattering as well as the diffuse scattering from a surface with a layer of particles on top of it.

\section*{Acknowledgement}
This work is supported by the Agence Nationale de la Recherche under project ANR-12-PRGE-0003-07

\appendix

\section{ \label{sec:microscopic}  Microscopic analysis of scattering from a mono-layer of particles in a homogeneous medium}


\begin{figure}[Hhbt]
\begin{center}
\includegraphics[trim = 0cm 0cm 0cm 0cm, clip=true, scale=0.5]{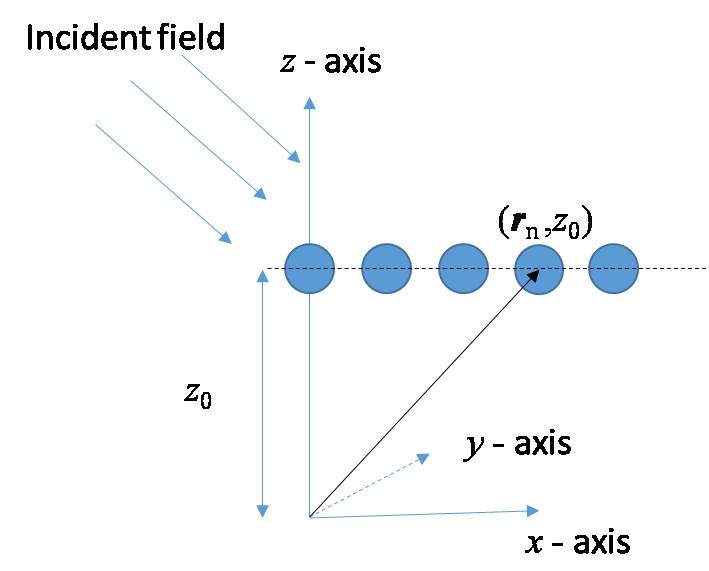}
\caption{Configuration for finding the scattered field at any point $(\boldsymbol{r}, z)$ from a layer of particle with centres located at $(\boldsymbol{r}_n, z_0)$ in a homogeneous medium of dielectric function $\epsilon_1$.  }
\label{layer}
\end{center}
\end{figure}
Here we consider a homogeneous medium containing a monolayer of a large number $N$ of identical spherical particles with center of the particles at ($\boldsymbol{r}_1,z_0$), ($\boldsymbol{r}_2,z_0$)...($\boldsymbol{r}_N,z_0$) as shown in Fig. \ref{layer}. We assume that the particles are uniformly distributed so that the medium is statistically homogeneous. Propagation of an electromagnetic wave in such a medium is characterized by the presence of a mean field $\langle \mathbf{E} (\mathbf{r}, z) \rangle$  corresponding to a plane wave in the statistically homogeneous medium and a fluctuating component corresponding to the fluctuating dielectric function \cite{durant2007light}. Thus when an incident field given by Eq. \ref{Inc} impinges on the layer of particles, the scattered field from the particles comprises of both a coherent component as well as a diffuse component (in contrast with the case of scattering by a single particle discussed in Sec. \ref{sec:single} with only a diffuse component). 

The mean field $\langle \mathbf{E} (\mathbf{r}, z) \rangle$  is given by:
\be
\label{mfield}
\langle \mathbf{E} (\mathbf{r}, z) \rangle = \int P(\boldsymbol{r}_1, \boldsymbol{r}_2, ..., \boldsymbol{r}_N) \, \, \mathbf{E} (\mathbf{r}, z)\, \, d^2 \boldsymbol{r}_1 \, \, d^2 \boldsymbol{r}_2  ...  d^2 \boldsymbol{r}_N
\ee
where, $P(\boldsymbol{r}_1, \boldsymbol{r}_2, ..., \boldsymbol{r}_N)$ is the probability density of finding particles at positions $(\boldsymbol{r}_1, z_0)$, $(\boldsymbol{r}_2, z_0)$,...,$(\boldsymbol{r}_N, z_0)$
and $\mathbf{E} (\mathbf{r}, z)$ is the field at position $(\mathbf{r}, z)$ given by:
\be
\label{eq:Esca}
\mathbf{E} (\mathbf{r}, z)   =   \sum_n \mathbf{E}_s^{(n)} (\mathbf{r} - \boldsymbol{r}_n, z-z_0)
\ee
where $\mathbf{E}_s^{(n)} (\mathbf{r} - \boldsymbol{r}_n, z-z_0)$ denotes the scattered field at $(\boldsymbol{r}, z)$ from a particle located at position $(\boldsymbol{r}_n, z_0)$. The scattered field from a particle is given by Eq. \ref{Fs12}, so that Eq. \ref{eq:Esca} can be written as:
\be
\label{total_sca_field}
\mathbf{E} (\mathbf{r}, z)   =   \sum_n  \int_{\mathbf{k_{||}^f}}  \frac{d^2\mathbf{k_{||}^f}}{(2 \pi)^2}  \mathbf{F}_n (\mathbf{k_{||}^f}, z-z_0) e^{i \mathbf{k_{||}^f}.( \mathbf{r}-\boldsymbol{r}_n)}.
\ee
The dipole moment of the particle located at ($\boldsymbol{r}_n, z_0$), $\mathbf{p}_0^{(n)}$, can be written as:
\be
\mathbf{p}_0^{(n)} = \varepsilon_0 \alpha_{\text{eff}} \mathbf{E}_{\text{ill}}^{(n)}
\ee
where $\mathbf{E}_{\text{ill}}^{(n)}$ is the field incident on the particle at position ($\boldsymbol{r}_n, z_0$) comprising of not just the incident field  $\mathbf{E}_i$  but also the scattered field at ($\boldsymbol{r}_n, z_0$)  from the other particles in the layer. Thus the scattered field from Eq. \ref{total_sca_field} reduces to:
\be
\label{total_sca_field2}
\mathbf{E} (\mathbf{r}, z)   =   \sum_n  \int_{\mathbf{k_{||}^f}}  \frac{d^2\mathbf{k_{||}^f}}{(2 \pi)^2}  \frac{i}{2} \alpha_{\text{eff}}  \, \varepsilon_1   \left( \frac{\omega}{c} \right)^2 \frac{e^{i{k}_{z1} (z-z_0)}}{k_{z1}} e^{i \mathbf{k_{||}^f}.( \mathbf{r}-\boldsymbol{r}_n)}  \Big[  \uv{s}\uv{s} +  \uv{p}_{1+}\uv{p}_{1+}  \Big] \cdot \mathbf{E}_{\text{ill}}^{(n)}
\ee
The mean field is thus given by:
\be
\label{total_sca_field3}
\langle \mathbf{E} (\mathbf{r}, z) \rangle   =    \int_{\mathbf{k_{||}^f}}  \frac{d^2\mathbf{k_{||}^f}}{(2 \pi)^2}  \frac{i}{2} \alpha_{\text{eff}} \, \varepsilon_1   \left( \frac{\omega}{c} \right)^2 \frac{e^{i{k}_{z1} (z-z_0)}}{k_{z1}} e^{i \mathbf{k_{||}^f}. \mathbf{r}} \Big\langle   \sum_n  \Big[  \uv{s}\uv{s} +  \uv{p}_{1+}\uv{p}_{1+}  \Big] \cdot \mathbf{E}_{\text{ill}}^{(n)} e^{-i \mathbf{k_{||}^f}. \boldsymbol{r}_n} \Big\rangle
\ee
For brevity we simplify Eq. \ref{total_sca_field3} for TE polarized waves. Similar arguments can be adopted for analyzing TM polarized waves. We first make an assumption that there is no correlation between the positions of the particles i.e., we write $P(\boldsymbol{r}_1, \boldsymbol{r}_2, ..., \boldsymbol{r}_N) = P_1(\boldsymbol{r}_1) P_2(\boldsymbol{r}_2)... P_N(\boldsymbol{r}_N)$ where  $P_i(\boldsymbol{r}_i)$ is the probability of finding a particle at position $(\boldsymbol{r}_i, z_0)$.   For non-correlation we have $P_i(\boldsymbol{r}_i) = 1/S$, $(i = 1,2,...N)$  where $S$ is the area in the $xy$ plane where the particles are located.  This assumption is valid in the dilute regime where the filling fraction of the particles given by $f = N \pi a^2/S \approx 0$ where $N$ is the total number of particles. For higher filling fractions correlation between the positions of the particles have to be taken into account as detailed in Ref. \cite{durant2007light}.
We also take
\be
\label{illum}
\mathbf{E}_{\text{ill}}^{(n)} = K \mathbf{E}_i (\boldsymbol{ r}, z)
 \ee
 where $K$ is an unknown quantity which accounts for the variation of the illuminated field from the incident field $\mathbf{E}_i (\boldsymbol{ r}, z)$ due to the scattered field. For an incident TE wave, taking the definition of mean-field from Eq. \ref{mfield} and substituting $\lim_{S \to \infty} \dfrac{1}{S} \int_S e^{i (\mathbf{k}_{\text{inc}}^{||}-\mathbf{k_{||}^f}). \boldsymbol{r}_n} d^2 \boldsymbol{r}_n =  \lim_{S \to \infty} \dfrac{(2 \pi)^2}{S} \delta(\mathbf{k}_{\text{inc}}^{||}-\mathbf{k_{||}^f}) $ where $\delta$ is the dirac delta function, Eq. \ref{total_sca_field3} reduces to:
\be
\label{total_sca_field4}
\langle \mathbf{E} (\mathbf{r}, z) \rangle   =   \lim_{S \to \infty} (N/S)\,\, K \frac{i}{2} \alpha_{\text{eff}} \, \varepsilon_1   \left( \frac{\omega}{c} \right)^2 \frac{1}{k_{z1}} e^{i \mathbf{k_{inc}}. \mathbf{r}} \,\,\uv{s}
\ee
%
%
%

The reflection coefficient of the coherently reflected light from the layer of particles is thus given by:
\be
\label{ref_coeff}
\hat{R}^{\text{TE}} = \dfrac{ \langle E (\mathbf{r}, z_0) \rangle}{E_i (\boldsymbol{ r}, z_0)} = \rho  \,  K \frac{i}{2} \alpha_{\text{eff}} \, \varepsilon_1   \left( \frac{\omega}{c} \right)^2 \frac{1}{k_{z1}}
\ee
where $\langle E (\mathbf{r}, z_0) \rangle$, and $E_i (\boldsymbol{ r}, z_0)$ are the complex scalar magnitudes of the vector quantities $\langle \mathbf{E} (\mathbf{r}, z_0) \rangle$, $\mathbf{E}_i (\boldsymbol{ r}, z_0)$  respectively and $\rho =  \lim_{S \to \infty}  \dfrac{ N}{S}  $ is the number of particles per unit area.
 To find the coherently transmitted beam below the layer of particles, thickness of which is less than the skin depth of the material, we take the field to be continuous across the layer of particles \cite{bauer1992optical}
i.e.,
\be
\label{T_TE_1}
\hat{T}^{\text{TE}} = (1 + \hat{R}^{\text{TE}})
\ee
where $\hat{T}^{\text{TE}}$ is the transmission coefficient of the incident TE wave.  The unknown factor $K$ can be determined by considering the field surrounding the layer of particles. Since we have
a field $(1 + \hat{R}^{\text{TE}})\mathbf{E}_i (\boldsymbol{ r}, z)$ surrounding the particles this should also be the illuminating field in Eq. \ref{illum}, i.e., we get $(1+\hat{R}^{\text{TE}}) = K$, which, on inserting in Eq. \ref{ref_coeff}, gives us:
\be
\label{R_TE_final}
\hat{R}^{\text{TE}} = \frac{ \rho \dfrac{i}{2} \left( \dfrac{\omega}{c} \right)^2 \dfrac{\varepsilon_1 \, \alpha_{\text{eff}}}{k_{z1}} }{1 - \rho \dfrac{i}{2} \left( \dfrac{\omega}{c} \right)^2 \dfrac{\varepsilon_1 \alpha_{\text{eff}}}{k_{z1}} }
\ee
and inserting  Eq. \ref{R_TE_final} in Eq. \ref{T_TE_1} we get:
\be
\label{T_TE_final}
\hat{T}^{\text{TE}} = \dfrac{1}{1 - \rho \dfrac{i}{2} \left( \dfrac{\omega}{c} \right)^2 \dfrac{\varepsilon_1 \, \alpha_{\text{eff}}}{k_{z1}} }
\ee
Eqs. \ref{R_TE_final} and \ref{T_TE_final} are equivalent to the coefficients obtained using the effective index model in Sec. \ref{sec:eff_index}.

\section{ \label{sec:anisotropy}  Fresnel reflection and transmission coefficients for TM polarization}
Here, we give the form of the Fresnel reflection and transmission coefficients to be used in Eqs. \ref{RthinFilm} and \ref{TthinFilm} when a TM polarized wave is incident on a layer of particles which is modeled as a thin-film using Maxwell-Garnett theory as described in Sec. \ref{sec:eff_index}. Since the dielectric property of the film, as given from Eq. \ref{dielectric_an}, has different components along $x$ and $z$ directions the expressions for $r$, $t^{(1)}$ and $t^{(2)}$ to be used in Eqs. \ref{RthinFilm} and \ref{TthinFilm} are given by \cite{lekner1987theory}:
\be
r = \frac{k_{z1}/\varepsilon_1 - k_{\text{eff}}/\varepsilon_{\text{eff},x}}{k_{z1}/\varepsilon_1 + k_{\text{eff}}/\varepsilon_{\text{eff},x} }
\ee
\be
t^{(1)} = \frac{2 k_{z1}/\varepsilon_1 }{k_{z1}/\varepsilon_1 + k_{\text{eff}}/\varepsilon_{\text{eff},x} } \times \sqrt{\frac{\varepsilon_1}{\varepsilon_{\text{eff},x}} }
\ee
\be
t^{(2)} = \frac{2 k_{\text{eff}}/\varepsilon_{\text{eff},x} }{k_{z1}/\varepsilon_1 + k_{\text{eff}}/\varepsilon_{\text{eff},x} }\times \sqrt{\frac{\varepsilon_{\text{eff},x}}{\varepsilon_1 } }
\ee

where
\begin{align}
k_{\text{eff}} = \sqrt{\varepsilon_{\text{eff},x} (\omega/c)^2  - \frac{\varepsilon_{\text{eff},x}}{\varepsilon_{\text{eff},z}} \mathbf{k}_{\text{inc}}^{||2} } 
\end{align}


\section*{References}
\begin{thebibliography}{10}
\expandafter\ifx\csname url\endcsname\relax
  \def\url#1{\texttt{#1}}\fi
\expandafter\ifx\csname urlprefix\endcsname\relax\def\urlprefix{URL }\fi
\expandafter\ifx\csname href\endcsname\relax
  \def\href#1#2{#2} \def\path#1{#1}\fi

\bibitem{schroder2011modeling}
S.~Schr{\"o}der, A.~Duparr{\'e}, L.~Coriand, A.~T{\"u}nnermann, D.~H. Penalver,
  J.~E. Harvey, Modeling of light scattering in different regimes of surface
  roughness, Optics express 19~(10) (2011) 9820--9835.

\bibitem{hermansson2003review}
P.~Hermansson, G.~Forssell, J.~Fagerstr{\"o}m, A review of models for
  scattering from rough surfaces, Link{\"o}ping: FOI-Swedish Defence Research
  Agency.

\bibitem{elfouhaily2004critical}
T.~M. Elfouhaily, C.-A. Gu{\'e}rin, et~al., A critical survey of approximate
  scattering wave theories from random rough surfaces, Waves in Random Media
  14~(4) (2004) R1--R40.

\bibitem{chandrasekhar2013radiative}
S.~Chandrasekhar, Radiative transfer, Courier Corporation, 2013.

\bibitem{thomas2002radiative}
G.~E. Thomas, K.~Stamnes, Radiative transfer in the atmosphere and ocean,
  Cambridge University Press, 2002.

\bibitem{mishchenko2002scattering}
M.~I. Mishchenko, L.~D. Travis, A.~A. Lacis, Scattering, absorption, and
  emission of light by small particles, Cambridge university press, 2002.

\bibitem{dahan2012enhanced}
N.~Dahan, J.-J. Greffet, Enhanced scattering and absorption due to the presence of
  a particle close to an interface, Optics express 20~(104) (2012) A530--A544.

\bibitem{draine1988discrete}
B.~T. Draine, The discrete-dipole approximation and its application to
  interstellar graphite grains, The Astrophysical Journal 333 (1988) 848--872.

\bibitem{carminati2006radiative}
R.~Carminati, J.-J. Greffet, C.~Henkel, J.~Vigoureux, Radiative and
  non-radiative decay of a single molecule close to a metallic nanoparticle,
  Optics Communications 261~(2) (2006) 368--375.

\bibitem{langlais2014cooperative}
M.~Langlais, J.-P. Hugonin, M.~Besbes, P.~Ben-Abdallah, Cooperative
  electromagnetic interactions between nanoparticles for solar energy
  harvesting, Optics Express 22~(103) (2014) A577--A588.

\bibitem{jylha2006modeling}
L.~Jylh{\"a}, I.~Kolmakov, S.~Maslovski, S.~Tretyakov, Modeling of isotropic
  backward-wave materials composed of resonant spheres, Journal of Applied
  Physics 99~(4) (2006) 043102.

\bibitem{sipe1987new}
J.~E. Sipe, New green-function formalism for surface optics, JOSA B 4~(4)
  (1987) 481--489.

\bibitem{bohren98a}
C.~F. Bohren, D.~R. Huffman, Absorption and Scattering of Light by Small
  Particles, Wiley-Interscience, 1998.

\bibitem{durant2007light}
S.~Durant, O.~Calvo-Perez, N.~Vukadinovic, J.-J. Greffet, Light scattering by a
  random distribution of particles embedded in absorbing media: diagrammatic
  expansion of the extinction coefficient, JOSA A 24~(9) (2007) 2943--2952.

\bibitem{chew95a}
W.~C. Chew, Waves and Fields in Inhomogeneous Media, IEEE Press, Piscataway,
  NJ, 1995.

\bibitem{born1999principles}
M.~Born, E.~Wolf, Principles of optics: electromagnetic theory of propagation,
  interference and diffraction of light, CUP Archive, 1999.

\bibitem{schmidt2012case}
K.~Schmidt, M.~A. Yurkin, M.~Kahnert, A case study on the reciprocity in light
  scattering computations, Optics express 20~(21) (2012) 23253--23274.

\bibitem{durant2007lightb}
S.~Durant, O.~Calvo-Perez, N.~Vukadinovic, J.-J. Greffet, Light scattering by a
  random distribution of particles embedded in absorbing media: full-wave monte
  carlo solutions of the extinction coefficient, JOSA A 24~(9) (2007)
  2953--2962.

\bibitem{bauer1992optical}
S.~Bauer, Optical properties of a metal film and its application as an infrared
  absorber and as a beam splitter, American journal of physics 60~(3) (1992)
  257--261.

\bibitem{lekner1987theory}
J.~Lekner, Theory of Reflection of Electromagnetic and Particle Waves, Vol.~3,
  Springer, 1987.

\end{thebibliography}
\end{document}